\documentclass[runningheads]{llncs}
\usepackage[T1]{fontenc}
\usepackage{graphicx}
\usepackage{algpseudocode}%
\usepackage{fancyhdr}

\usepackage{algorithm}
\usepackage{algpseudocode}

\usepackage{amsmath}
\usepackage{cite}
\usepackage{hyperref}
\usepackage{amsmath} 
\usepackage{amssymb}
\usepackage{bm}
\usepackage{booktabs}
\usepackage{makecell}
\usepackage{array,color}
\usepackage{array}
\usepackage[table]{xcolor}
\usepackage{multirow}

\usepackage[misc]{ifsym}

\usepackage{multirow}  
\usepackage{booktabs} 
\usepackage{threeparttable}
\usepackage{bbding}
\usepackage{ulem}
\begin{document}
\bibliographystyle{splncs04}

\title{Topology-Aware Exploration of Circle of Willis for CTA and MRA: Segmentation, Detection, and Classification}
\titlerunning{Circle of Willis (CoW): Segmentaion, Detection, and Classification}
\author{
Minghui Zhang\inst{1,2} \and 
Xin You \inst{1,2}
Hanxiao Zhang\inst{1}  \and 
Yun Gu\inst{1,2}\textsuperscript{(\Letter)}
}

\authorrunning{M. Zhang et al.}
\institute{Institute of Medical Robotics, Shanghai Jiao Tong University, Shanghai, China
\email{
   \{minghuizhang, geron762\}@sjtu.edu.cn}\\
\and
Department of Automation, Shanghai Jiao Tong University, Shanghai, China
}
\maketitle              

\begin{abstract}
The Circle of Willis (CoW) vessels is critical to connecting major circulations of the brain. 
The topology of the vascular structure is clinical significance to evaluate the risk, severity of the neuro-vascular 
diseases. The CoW has two representative angiographic imaging modalities, computed tomography angiography (CTA) 
and magnetic resonance angiography (MRA). TopCow24 provided 125 paired CTA-MRA dataset for the 
analysis of CoW. To explore both CTA and MRA images in a unified framework to learn the 
inherent topology of Cow, we construct the universal dataset via independent intensity preprocess, 
followed by joint resampling and normarlization. Then, we utilize the topology-aware loss to enhance the 
topology completeness of the CoW and the discrimination between different classes. A complementary topology-aware 
refinement is further conducted to enhance the connectivity within the same class. Our method was evaluated on 
all the three tasks and two modalities, achieving competitive results. In the final test phase of TopCow24 Challenge, 
we achieved the second place in the CTA-Seg-Task, the third palce in the CTA-Box-Task, the first place in the CTA-Edg-Task, 
the second place in the MRA-Seg-Task, the third palce in the MRA-Box-Task, the second place in the MRA-Edg-Task.
\keywords{Circle of Willis \and Connectivity-Aware Loss \and Topological Refinement}
\end{abstract}

\section{Introduction}
The Circle of Willis (CoW) vessels is critical to connecting the circulations of the brain, 
including the anterior and posterior, and left and right cerebral hemispheres \cite{price2014osborn}. The CTA and 
MRA are two main modalities to dignose the risk, severity of the neuro-vascular diseases. However, characterizing 
the anatomy of the CoW is time-consuming and relies on the knowledge of experts. Furthermore, the paired CTA and MRA 
image for the same patients are rare. TopCow23 and TopCoW24 \cite{yang2023benchmarking} has provided the paired CTA 
and MRA images with multi-class CoW labels for research. The paired scans of CTA and MRA are significant to enlarge 
the dataset, however, explore these data directly in a unified framework may be problematic since the existence of the 
modality difference. In addition, the unbalanced class distribution across the classes adds the difficulty in 
preserving the correct topology. Furthermore, the small breakage could happen, similar to other tubular structures \cite{zhang2023multi}, 
which is harmful to the overall topology. To tackle the aforementioned challenges, we propose three key components 
to alleviate them, respectively. First, the universal dataset is constructed for the modality-agnostic model optimization. 
Second, in TopCow23 challenges, several methods employed the centerline or skeleton for topological objects of interest \cite{kirchhoff2024skeleton,shi2024centerline,shit2021cldice, shi2023nextou}. 
The improved connectivity results demonstrated the effectiveness of these methods.
We also design the topology-aware loss to enhance the completeness of the CoW and increase the discrimination between different classes. 
The topology-aware loss dynamically adjusts the importance of the most center part of each class. This can implicitly enhance the distinction between classes.
Third, A complementary topology-aware refinement is further conducted to enhance the connectivity within the same class. 
Our method was evaluated on all the three tasks and two modalities, achieving competitive results.  

\begin{figure}[t]
\centering
\includegraphics[width=1.0\linewidth]{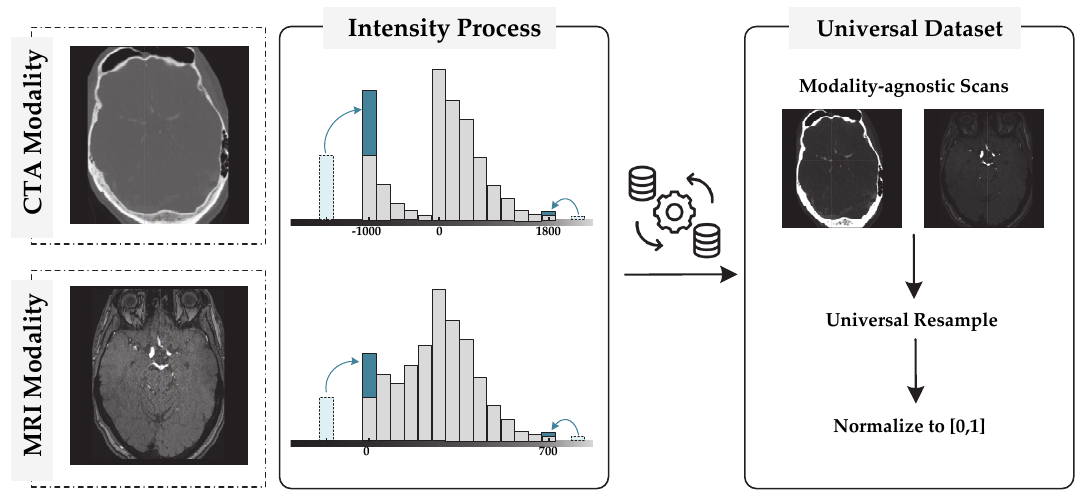}
\caption{The process to acquire the universal dataset from the CTA and MRI scans. $[-1000,1800]$ is used to 
truncate the CTA scans. As for the MRI scans, we chose $[0,700]$. 
After independently processing the intensity truncation, the modality-agnostic scans were constructed. These modality-agnostic scans are then universally resampled to the same resolution. 
Finally, we normalize the intensity of each case to $[0, 1]$. }
\label{fig:preprocess}
\end{figure}

\section{Method}
\subsection{Universal Dataset Construction} 
The organizers had provided the paired 125 CTA-MRI scans in the dataset. Despite the paired scans have not been rigidlly registered, we aggregate the scans both from the CTA domain and MRI domain 
to construct a universal dataset, similar to several teams performed in TopCow23\cite{yang2023benchmarking}. The intensity distribution exists significant difference between the CTA domain and MRI 
domain. The standard nnUNet\cite{isensee2021nnu} workflow supports the multi-modality inputs that preprocessed by independent intensity normalization strategy, while the provided paired CTA-MRI scans 
do not share the same Circle of Willis (CoW) label. In this situation, nnUNet may tend to process these scans separately for each modality. The topological structures of the CoW can be partially 
corresponded within the different modality of scans that belong to the same patient. Hence, our hypothesis is that the appropriation expansion of the dataset could be beneficial to the learning of 
the inherent toplogy of CoW. As seen in Fig.\ref{fig:preprocess}, different settings of the truncation values are applied for CTA and MRI scans, respectively. Concretely, we used $[-1000,1800]$ to 
truncate the CTA scans. As for the MRI scans, we chose $[0,700]$. Different from the lung/mediastinum window, we do not find the standard HU window to preprocess the CoW. Therefore, the chosen 
truncation values were obtained from the dataset information. We determined the intensity range based on the majority of voxel intensity that belongs to the CoW, meanwhile surpassing the abnormal value. 
After independently processing the intensity truncation, we constructed the modality-agnostic scans. These modality-agnostic scans are then universally resampled to the same resolution. 
Due to the modality-agnostic attribute, We do not perform Z-score normalization for MRI scans, nor do we apply 0.5\% and 99.5\% percentile clipping for CT scans. 
We simply normalize the intensity of each case to $[0, 1]$. 

\begin{figure}[t]
\centering
\includegraphics[width=0.95\linewidth]{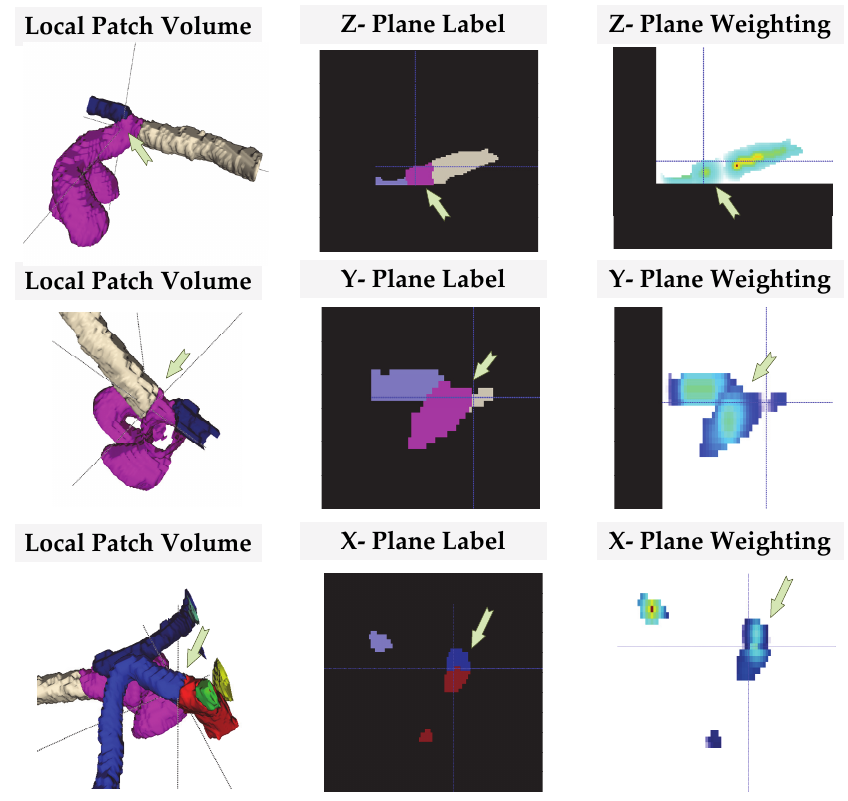}
\caption{The visualization of the dynamic weights of the connectivity-aware loss for multi-class vessels. The weighting is 
conducted on-the-fly based on the patch volume and independently calculated for each class.}
\label{fig:CAL}
\end{figure}

\subsection{Topology-aware Loss} 
Base on the universal dataset that enlarge the dataset and alleviate the modality difference, we further developed the 
connectivity-aware loss (CAL) based on \cite{zhang2023towards} into the nnUNet architecture to further learning the inherent topological structures behind the images. 
Specifically, The multi-class segmentation of Cow suffers from the class imbalance problem. The binarized CoW still only occupies a small portion of the scans, compared with the background. 
Moreover, the class imbalance between foregrouds is also significant in the multi-class CoW. For example, the class of BA, R-PCA, L-PCA can occupy more voxels than the R-Pcom and L-Pcom in the 
posterior part. However, all the five classes contribute significantly to the topological correctness for the diagnosis of the variants of the CoW.
Hence, CAL first includes the overlap-based optimization, we used the Dice with Focal loss for the basic performance of the multi-class segmentation task. 
Further, as guided in \cite{maier2024metrics}, only overlap-based optimization may not sufficient for the complex shapes, especially when the segmentation performance are evaluated by 
topology-related metrics. Therefore, the CAL is proposed to enhance the completeness of the multi-class segmentation of CoW. 
The difficulty of improving the completeness of the multi-class segmentation of CoW can be attributed to
within-class distribution imbalance. Large classes of Cow occupy themajority of foreground voxels, and such imbalanced distribution 
affects the data-driven deep learning methods, which may lead to poor performance on the peripheral small classes of CoW. To
enhance the completeness for small classes , we design the following objective function:

\begin{gather}
\begin{split}
    L_{c} = (-\frac{2\sum\nolimits_{\forall i}\mathbf{\hat{y}}_{i}^{c}\mathbf{y}_{i}^{c}}{\sum\nolimits_{\forall i}(\mathbf{\hat{y}}_{i}^{c}+\mathbf{y}_{i}^{c})} -\frac{1}{|I|}(\sum\limits_{\forall i}(1-\mathbf{\hat{y}}_{i}^{c})^2log(\mathbf{\hat{y}}_{i}^{c}))) \\
    + (1 - \frac{\sum_{i = 1}^{N} \mathbf{\hat{y}}_{i}^{c}\mathbf{y}_{i}^{c}}{\alpha_{t}\sum_{i}\mathbf{\hat{y}}_{i}^{c} + \beta_{t}\sum_{i}\mathbf{y}_{i}^{c}}) + (\sum_{i = 1}^{N} \gamma_{i}^{c} \mathrm{CE}(\mathbf{\hat{y}}_{i}^{c}, \mathbf{y}_{i}^{c})).
\end{split}
\end{gather} 

\begin{gather}
\begin{split}
    L_{total} = \frac{1}{|C|} \sum\nolimits_{c} L_{c},
\end{split}
\end{gather} 

where $\alpha_{t} + \beta_{t} = 1$. Both $\alpha_{t}$ and $\beta_{t}$ are hyper-parameters to balance the recall and sensitivity of segmentation. In our experiments, we chose $\alpha_{t}$ = 0.2 and $\beta_{t}$ = 0.8. 
$\gamma$ is the distance-based weighting map. $c$ denote the class of each class of CoW, in total thirteen classes are considered.
The weight of each voxel depends on the Euclidean distance to the centerline, which is defined as:
\begin{align}
\gamma_{i}^{c}=\left\{\begin{array}{ll} 
- \lambda_{\mathrm{fg}}\mathrm{log}(\frac{\mathrm{dc}_{i}}{\mathrm{dc}_{max}} + \varepsilon ),& \mathbf{y}_{i}^{c} = 1,\\
1,& \mathbf{y}_{i}^{c} = 0, \label{Loss_Weight-CE_alpha}
\end{array}\right.
\end{align}
where $\lambda_{\mathrm{fg}}$ is set to 20, and $\varepsilon$ is set to 0.01. $\mathrm{dc}_{i}$ and $\mathrm{dc}_{max}$ represent the distance from the certain 
voxel to the centerline and the max distance of one class, respectively. The on-the-fly topology-aware optimization is demonstrated in Fig.\ref{fig:CAL}, each class 
is re-weighted independently within the patch volume. The $\varepsilon$ can control the importance degree of the most center part of each class. This can implicitly 
enhance the distinction between classes. The $\lambda_{\mathrm{fg}}$ could emphasize the importance of the overall foreground.

\begin{figure}[t]
\centering
\includegraphics[width=1.0\linewidth]{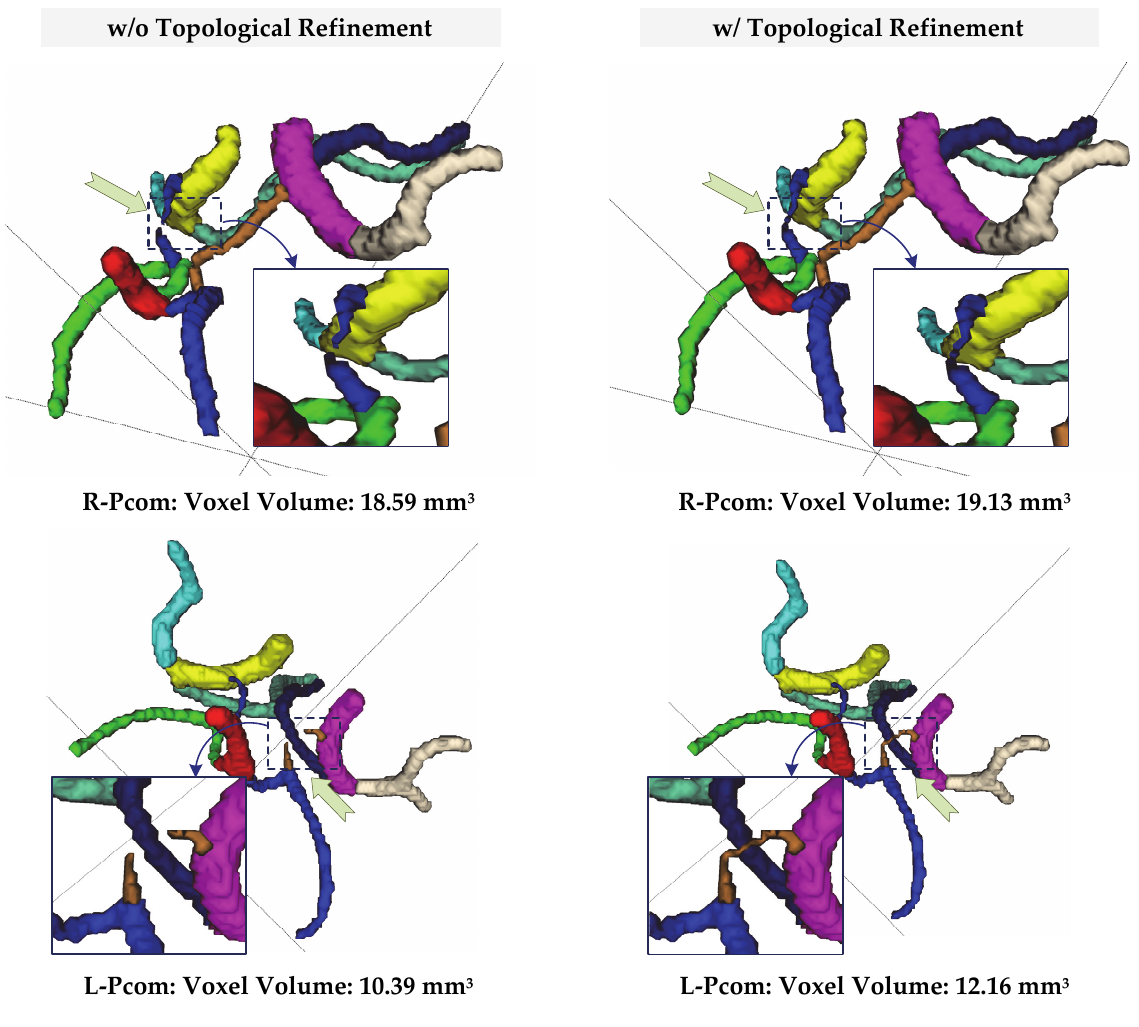}
\caption{The qualitative results of the topology-aware refinement.}
\label{fig:refinement}
\end{figure}

\begin{algorithm}[t]
    \label{algo:refinement}
    \caption{Topological Refinement for Partial CoW classes}
    \begin{algorithmic}[1] 
        \State \textbf{INPUT}: $Pred_{cls}$ $\gets$ initial prediction of the certain class
        \State \textbf{OUTPUT}: $Pred_{post}$ $\gets$ initial prediction of the certain class
        \State $T_{com}$ $\gets$ threshold for the meaningful component
        \State $T_{dis}$ $\gets$ threshold for the maximum distance for repairing
        \State Components, Num = Func-component-analysis($Pred_{cls}$)
        \If{$Num == 1$}
            \State \Return $Pred_{cls}$
        \ElsIf{$Num == 2$}
            \State Component-1, Component-2 $\gets$ Components
            \If{Component-1 \& Component-2 \textit{satisfy} $T_{com}$}
                \State Find the paired closet endpoints [endpoint-1, endpoint-2]
                    \If{Dist(endpoint-1, endpoint-2)  \textit{satisfy} $T_{dis}$}
                        \State  RepairComponent  $\gets$ dilation-interpolate(endpoint-1, endpoint-2) 
                        \State  $Pred_{post}$ $\gets$ Components + Repair Component 
                        \State \Return  $Pred_{post}$
                    \Else 
                        \State \Return largest-connected-component($Pred_{cls}$)
                    \EndIf
            \ElsIf{not Component-1 | Component-2 \textit{satisfy} $T_{com}$}
                \State \Return Zeros-like($Pred_{cls}$)
            \Else 
                \State \Return largest-connected-component($Pred_{cls}$)
            \EndIf
        
        \Else
            \State Extract the two largest connected components
            \State \Return Subprocess($Num == 2$)
            \EndIf
    \end{algorithmic}
\end{algorithm}

\subsection{Topology-aware Refinement} 
As reported in the TopCow23 benchmark paper\cite{yang2023benchmarking}, the representative CoW variants of the anterior and posterior parts have been reported. To guarantee the 
topological correctness of these part, we propose the simple-yet-effective topology-aware refinement module as the postprocess procedure. The full pseudo algorithm can be seen 
in Algorithm.\textbf{1}. Basiclly, it is a rule-based method that conforms to the topological prior of the CoW structure. 
For example, when two disconnected significant components of the 
class of R-Pcom are inferred by the model with high confidence. One of the two components connects to R-PCA, another connects to R-ICA. 
The class of R-Pcom should be recontructed as either the sigle connected component or inexistence.
Under this situation, the post-repair procedure should be considered. We first determine the paired closet endpoints based on their skeletons, then we interpolate the endpoints for the spine interpolation usage. 
The generated points are then slightly dilated to be connected with the initial prediction. The effectiveness of this component could be found in the Fig.\ref{fig:refinement}. We exhibited two classes as 
examples, the R-Pcom and the L-Pcom, the refinement procedure affects negligible voxel volumes, however, it could enhance the topological correctness of the topology. The key parameters are the 
$T_{com}$ and $T_{dis}$. $T_{com}$ measures the lower limit of the meaningful component. $T_{dis}$ determines the upper limit of the discontinuity distance for repairing. 
These parameters could be determined from the statistics of each class by the provided ground-truth of the classes.

\section{Experimental Analysis}
\subsection{Implementation Detail}
The universal dataset was used in all the tracks. nnUNet\cite{isensee2021nnu} was chosen as the backbone structure. We do not observe the significant gain from the 
model ensemble of the multiple fold result, hence, we chose to use the single fold model for the predition. The single fold model could also guarantee the time efficiency.
The provided RoIs were not used to crop the RoIs for all the tracks, all the models were trained on the whole image, since the discrimination between the foreground and background is also beneficial to 
the learning of complex shapes. All scans were uniformly resampled to the spacing (order of the z,y,x) of [0.6$mm$,0.3525$mm$,0.3525$mm$]. 
The patch-size of 80 $\times$ 192 $\times$ 160 and batchsize of 2 were used in the training. The augmentation within the nnUNet workflow was used except the 
mirror operation to avoid the confusion of the symmetrical classes. $\lambda_{\mathrm{fg}}$ is set to 20, and $\varepsilon$ is set to 0.01 in our Topology-aware Loss. 
These two parameters were determined on our local split validation set, and also tested another set of $\lambda_{\mathrm{fg}} = 10, \varepsilon = 0.05 $ on the CTA multi-class 
segmentation on the online validation phase.For simplicity, we used $T_{com}=20$, $T_{dis}=10$ in the experiments while we had known it should be better to obtain the statistical priors from the dataset fingerprint. 

\subsection{Task-1: Multi-class Segmentation of CoW (CTA and MRI)}
The segmentation task is voxel-level multi-class CoW vessel segmentation.
The vessel components of the CoW annotated are left and right internal carotid artery (ICA), left and right anterior cerebral artery (ACA), 
left and right middle cerebral artery (MCA), anterior communicating artery (Acom), left and right posterior communicating artery (Pcom),
left and right posterior cerebral artery (PCA), and basilar artery (BA). The Tab.\ref{tab:seg} reported the results achieved by our method on both the 
validation and test phase for the CTA and MRI tracks. All results could be found online\footnote{\url{https://topcow24.grand-challenge.org}}. The proposed 
CAL could guarantee the topological completeness to achieve the high class average dice, cldice. The emphasis of the central parts and explore on the small 
vessels show the effectiveness in achieving acceptable Class Average HD95, and Class Average B0. The topological refinement module could enhance the 
topological-related metrics meanwhile maintain the overlap-wised metrics.

\makeatletter
\def\hlinew#1{%
\noalign{\ifnum0=`}\fi\hrule \@height #1 \futurelet
\reserved@a\@xhline}
\makeatother
\begin{table}[h]
\renewcommand\arraystretch{1.4}
\caption{Quantitative results of the multi-class segmentation on both CTA and MRI.}\label{tab:seg}
\centering
\scalebox{0.7}{
\begin{tabular}{@{}l|ccc|cc@{}}
\toprule
\multicolumn{1}{c|}{\multirow{2}{*}{\textbf{Metrics}}}       & \multicolumn{3}{c|}{\textbf{Validation Phase}}   & \multicolumn{2}{c}{\textbf{Test Phase}} \\ \cmidrule(l){2-6} 
\multicolumn{1}{c|}{}                               & CTA-1   & CTA-2     & MRI       & CTA            & MR            \\ \midrule
Class Average Dice Per Case                         & 0.85 $\pm$ 0.05 & 0.86 $\pm$ 0.05 & 0.90 $\pm$ 0.04 & 0.87 $\pm$ 0.06      & 0.88 $\pm$ 0.07     \\
clDice                                              & 0.99 $\pm$ 0.01 & 0.99 $\pm$ 0.01 & 0.98 $\pm$ 0.02 & 0.99 $\pm$ 0.02      & 0.98 $\pm$ 0.02     \\
Class Average B0 Error Per Case                     & 0.09 $\pm$ 0.06 & 0.02 $\pm$ 0.04 & 0.02 $\pm$ 0.04 & 0.06 $\pm$ 0.09      & 0.09 $\pm$ 0.13     \\
Class Average HD95 Per Case                         & 0.95 $\pm$ 0.30 & 0.82 $\pm$ 0.29       & 0.83 $\pm$ 0.47 & 2.88 $\pm$ 3.71      & 2.57 $\pm$ 3.66     \\
Average F1 Grp2                                     & 0.75 $\pm$ 0.43 & 0.75 $\pm$ 0.43       & 0.71 $\pm$ 0.42 & 0.86 $\pm$ 0.08      & 0.91 $\pm$ 0.02     \\
Anterior Variant Balanced Graph Classifcation Acc   & 1.00      & 1.00            & 1.00      & 0.91           & 0.91          \\
Posterior Variant Balanced Graph Classification Acc & 0.78      & 1.00            & 0.75      & 0.63           & 0.69          \\
Anterior Variant Balanced Topology Match Rate       & 0.88      & 0.88            & 0.88      & 0.57           & 0.52          \\
Posterior Variant Balanced Topology Match Rate      & 0.11      & 0.89            & 0.50      & 0.50           & 0.49          \\ \bottomrule
\end{tabular}}
\end{table}

\makeatletter
\def\hlinew#1{%
\noalign{\ifnum0=`}\fi\hrule \@height #1 \futurelet
\reserved@a\@xhline}
\makeatother
\begin{table}[h]
\renewcommand\arraystretch{1.4}
\caption{Quantitative results of the CoW RoI detection on both CTA and MRI.}\label{tab:det}
\centering
\scalebox{0.8}{
\begin{tabular}{@{}c|cc|cc@{}}
\toprule
\multirow{2}{*}{\textbf{Metrics}} & \multicolumn{2}{c|}{\textbf{Validation Phase}} & \multicolumn{2}{c}{\textbf{Test Phase}} \\ \cmidrule(l){2-5} 
                            & CTA               & MRI               & CTA            & MRI           \\ \midrule
Boundary IoU             & 0.62 $\pm$ 0.07         & 0.66 $\pm$ 0.11         & 0.62 $\pm$ 0.10      & 0.67 $\pm$ 0.11     \\
IoU                      & 0.74 $\pm$ 0.05         & 0.77 $\pm$ 0.08         & 0.74 $\pm$ 0.08      & 0.77 $\pm$ 0.10     \\ \bottomrule
\end{tabular}}
\end{table}

\makeatletter
\def\hlinew#1{%
\noalign{\ifnum0=`}\fi\hrule \@height #1 \futurelet
\reserved@a\@xhline}
\makeatother
\begin{table}[h]
\renewcommand\arraystretch{1.4}
\caption{Quantitative results of the classification of the anterior and posterior parts of the CoW graph on both CTA and MRI.}\label{tab:cls}
\centering
\scalebox{0.85}{
\begin{tabular}{@{}c|cc|cc@{}}
\toprule
\multirow{2}{*}{\textbf{Metrics}} & \multicolumn{2}{c|}{\textbf{Validation Phase}} & \multicolumn{2}{c}{\textbf{Test Phase}} \\ \cmidrule(l){2-5} 
                            & CTA               & MRI               & CTA            & MRI           \\ \midrule
Anterior Variant Balanced Graph Classifcation Acc             & 1.00        & 1.00      & 0.87     & 0.86    \\
Posterior Variant Balanced Graph Classification Acc                      & 1.00         & 0.75         & 0.64     & 0.69 \\ \bottomrule
\end{tabular}}
\end{table}

\subsection{Task-2: CoW RoI Detection (CTA and MRI)}
The Task-2 belongs to the object detection. The provided labels show that the regions needs to be detected are the RoIs of the whole CoW. 
It need to locate a 3D bounding box for the CoW RoI. The CoW RoI is defined as the 3D bounding box containing the volume required for the diagnosis of the CoW variant.
We observed that it is hard to directly determine the RoI from the whole prediction results. We do not further perform the RoI regression head. Instead, 
We replace the ground-truth of the whole CoW with the RoI CoW, and we still use the whole image to retrain a segmentation head. Then the detection task was 
transformed to the segmentation task. After acquiring the RoI voxel-wise prediction, we extract the class-wise largest connected component to obtain the final 
tight bounding box. Tab.\ref{tab:det} reported our detection results by the Boundary IoU and IoU for CTA and MRI scans.

\subsection{Task-3: CoW graph Classification (CTA and MRI)}
The classification task is to classify the anterior and posterior parts of the CoW graph in the form of two edge lists per image. 
For each image, the edges of the CoW graph indicates the presence of edges (0: absent, 1: present). The anterior part and the posterior part are 
taken into consideration. 4 edges for anterior part, roughly from left to right: L-A1, Acom, 3rd-A2, and R-A1. 4 edges for posterior part, roughly from left to right:
L-Pcom, L-P1, R-P1, and R-Pcom. Similar to Task-2, we do not further train the classification head for this task. We directly deduce the CoW graph classification based 
on the prediction results from the Task-1. The results are reported in the Tab.\ref{tab:cls}.

\section{Discussion and Future Directions}
\noindent\textbf{Data Integration:} It is observed that the truncated values can not handle all foregrounds, especially the extreme intensity 
occured. Few wrong prediction of voxels caused by the abnormal value can lead to the obvious topological errors. Hence, 
We will explore better integration of different modalities.\\
\noindent\textbf{Inter-class Penalty:} The CAL primarily improve the weight of the center part of each class, however, the penalty on the misclassification has not been explored fully. 
Therefore, we schedule to add the Inter-class misclassification penalty and purse consistency improvement among all the classes in the future research.\\

\noindent\textbf{Acknowledgement.}
This work is supported in part by the Open Funding of Zhejiang
Laboratory under Grant 2021KH0AB03, in part by the Shanghai Sailing Program
under Grant 20YF1420800, and in part by NSFC under Grant 62003208, and in part
by Shanghai Municipal of Science and Technology Project, under Grant 20JC1419500
and Grant 20DZ2220400.

\bibliography{paper402.bib}
\end{document}